\documentstyle[aps,epsf,preprint,prb]{revtex}
\def\DESepsf(#1 width #2){\epsfxsize=#2 \efsfbox{#1} \vspace*{0.07in}}
\def \DESepsf(#1 width #2){\bf #1  here: just uncomment the macro.}
\begin{document}

\bibliographystyle{unsrt}
\draft

\title{Observation of Individual Josephson Vortices in YBCO Bicrystal
Grain-Boundary Junctions}

\author{H. Xin,$^{\dag \Diamond *}$ D. E. Oates,$^{\dag
\Diamond}$ S. Sridhar,$^{\S}$
 G. Dresselhaus,$^{\dag *}$ 
 M.~S.~Dresselhaus$^{\dag}$}
\address{$^{\dag}$ Department of Physics, Massachusetts Institute of Technology, Cambridge, MA 02139\\
$^{\Diamond}$ Lincoln Laboratory,  Lexington, MA 02420\\
$^{\S}$ Department of Physics, Northeastern University, Boston, MA 02115\\}
\address{$^*$AFRL, Hanscom AFB, Bedford, MA 01731\\}

\date{\today}
\pagebreak
\maketitle
\begin{abstract}
The response of YBCO bicrystal grain-boundary junctions to small dc
magnetic fields (0 - 10 \,Oe) has been probed with a low-power
microwave (rf) signal of 4.4 \,GHz in a microwave-resonator
setup. Peaks in the microwave loss at certain dc magnetic fields are
observed that result from individual Josephson vortices penetrating into the
grain-boundary junctions under study. The system is modeled as a long
Josephson junction described by the sine-Gordon equation with the
appropriate boundary conditions. Excellent quantitative agreement
between the experimental data and the model has been
obtained. Hysteresis effect of dc magnetic field is also studied and
the results of measurement and calculation are compared.

\end{abstract}
Grain-boundary junctions in high-temperature-superconducting (HTS)
thin films have been studied
extensively\cite{oates96,hein96,wosik96,cowie99,booth99,habib98a,habib98b,nguyen93,hump93}
due to their importance in both device applications and fundamental
physics. The dynamics of Josephson vortices in Josephson junctions,
which are analogous to the Abrikosov vortices in type II superconductors,
\cite{kulik67,goldman67} is a very interesting subject, because of the
mesoscopic physics involved. Much effort has been devoted to understanding
the collective behavior of Josephson vortices in situations such as
flux-flow devices.\cite{shenbook} The study of vortex physics on the
mesoscopic scale has in general been hampered by the lack of suitable
experimental probes. \cite{bolle99} It has also been known that for
long junctions (junctions with a dimension longer than the Josephson
penetration depth), Josephson vortices induced by the rf magnetic field 
cause nonlinear microwave losses which can severely limit the
applicability of HTS materials in wireless communication
applications.\cite{habib98a,oates99,clem97} Previous experiments and
modeling have yielded qualitative agreement.\cite{habib98a,oates99} To
further quantitatively understand the effect of the Josephson vortex
dynamics on the microwave losses of HTS materials, it is important to
study the influence of the Josephson vortices generated by dc magnetic
fields. In this scenario, the vortex dynamics should be easier to probe,
and an external dc magnetic field can
emulate the physical situations of trapped
flux or the earth's ambient magnetic field.       

We report the observation of individual Josephson vortices generated
by an external dc magnetic field and how these events affect microwave
losses. We probe the Josephson vortex dynamics by using a small rf
signal of 4.4 \,GHz in a microwave-resonator setup that includes a
bicrystal grain-boundary junction. Each Josephson vortex entering the
junction is manifested by a sharp peak in the microwave resistance,
which is measured by our experimental setup. The second-order 
nonlinear sine-Gordon equation, which determines the dynamics of a
long junction in the presence of dc and rf magnetic fields, is solved
numerically. The measured and calculated results agree quantitatively
as the dc field is increasing, and we are able to identify the series
of sharp peaks observed in the microwave loss with the first several
Josephson vortices penetrating into the grain-boundary
junctions. Hysteresis effects upon decreasing the dc magnetic field
have also been measured and calculated. Differences between calculated
and measured losses in decreasing field will be discussed. 

The junctions used in this study were fabricated from 140-\,nm-thick,
epitaxial, c-axis-oriented, YBCO films deposited by laser ablation on
1\,-cm by 1\,-cm r-plane (1012) sapphire
bicrystal substrates with a $24^{\circ}$-misorientation angle.\cite{nist96}
To characterize the microwave properties of the junction, we have used
a microstrip-resonator configuration that allows us to distinguish the
effects of the junction from those of the rest of the film. The
resonator was patterned such that the junction is positioned at the
midpoint of the microstrip, spanning the entire width of 150\,$\mu$m,
as shown in Fig.~\ref{fig1}. The resonance frequency $f_1$ of the fundamental
mode is 4.4\,GHz with overtone resonant modes at $f_n$ $\approx$
$n$$f_1$ where $n$ is an integer. At resonance, the fundamental 
mode is a half-wavelength standing wave with a current maximum at the
midpoint of the resonator line, where the fabricated junction is
positioned. In contrast, the $n=2$ mode is a full wavelength with a
current node at the position of the junction. Therefore, by comparing the
measured results of these two modes, we can separate the properties of
the engineered grain-boundary junction from those of the remainder of the
superconducting film. This resonator technique has previously been
used to measure the microwave power-handling properties of the engineered 
junction in zero dc magnetic field.\cite{habib98a} 

In this work, we have studied the characteristics of the junctions in
small dc magnetic fields by injecting low-power rf input signals (pW) at
the fundamental and the first overtone (4.4 and 8.7\,GHz) of the
resonator. The device was cooled in a magnetic field smaller than
0.01\,Oe. The quality factor $Q_0$, which is proportional to the inverse
of the microwave resistance, was measured for dc magnetic fields
ranging from 0 to 10\,Oe with a step size as small as 0.01\,Oe, at 
temperatures ranging from 5 to 75\,K. As expected, the measurements of
the $n=2$ mode showed no observable dc magnetic field dependence,
since in this mode, the engineered junction does not contribute to the
microwave loss, and the applied magnetic field $H_{dc}$ is too small
to affect the rest
of the film. However, in the measurements of the $n=1$ mode, we
observe an abrupt decrease in $Q_0$ at certain narrow ranges of dc
magnetic fields, followed by a recovery to the zero-field value. This
pattern is followed for all of the measured temperatures. Experimental
data, plotted as $1/{Q_0}$ (proportional to the microwave resistance)
versus dc magnetic field at various temperatures, are shown in
Fig.~\ref{fig2}. As described below, we interpret the observed peaks
in $1/{Q_0}$ as single Josephson vortices penetrating into the
junction. Furthermore, at each temperature, the data show a threshold
field below which the microwave loss is low, and at which the first
peak in the microwave loss is observed. The threshold field can be
identified as the critical Josephson field $H_{cJ}$. Also, notice that
the higher the temperature, the noisier the data becomes, indicating that
thermal fluctuations may cause nucleation and annihilation of
Josephson vortices. The noise is most apparent in the 75\,K data. Two
$24^{\circ}$ junction samples have been measured and almost identical
behavior was observed.


The length $L$ of the grain-boundary junction is 150\,$\mu$m, which is
much greater than the Josephson penetration depth $\lambda_J$ given
by,\cite{tinkbook}  
\begin{equation}
{\lambda}_{J} = \sqrt{\frac{{\Phi}_{0}}{2{\pi}{\mu}_0{J_c}(2\lambda_L+d)}},
\end{equation}
where $\Phi_0$ is the flux quantum,
$\Phi_0$=$h/2e$=$2.07\times10^{-15}\,Wb$, $\lambda_L$
$\approx$\,0.2\,$\mu$m is the London penetration depth of 
the film, and $d$ is the physical grain-boundary interlayer thickness,
which is negligible compared with $\lambda_L$. 
For a typical $J_c$ of
$10^2$ to $10^4$ A/cm$^2$, 
$\lambda_J \ll L$ and the long-junction
regime applies.\cite{tinkbook} For this situation, $H_{cJ}$ is
given by
\begin{equation} 
H_{cJ}=2{\mu}_{0}J_{c}{\lambda_J}{\propto}{\sqrt{{J_c}/{\lambda_L}}}\,.
\label{eq1}
\end{equation}
Above $H_{cJ}$, a Meissner state of the junction is not possible, and quantized
flux in the form of Josephson vortices starts to penetrate into the
junction from its edges. 
 
The dynamics of a long-junction system are governed by the sine-Gordon
equation,\cite{barone,portis92}
\begin{equation}
{\lambda_J^{2}}\frac{{\partial}^{2}\phi(x,t)}{{\partial}x^{2}} =
\sin{\phi(x,t)} + {\tau}_{J}\frac{{\partial}{\phi}}{{\partial}t},
\label{eq2}
\end{equation}
where $\phi(x,t)$ is the gauge-invariant phase difference of the
superconducting wave function across the junction
${\tau_J}={\Phi}_{0}/2{\pi}d{J_c}{\rho}_{n}$, with $\rho_n$ being the
normal leakage resistivity of the junction. The capacitive
term is omitted for our case of an overdamped junction.\cite{habib98a}
We have solved Eq.~(\ref{eq2}) numerically with boundary conditions at
the junction edges that include both the dc and microwave magnetic
field. Similar treatments have been reported by other
authors.\cite{clem97,sridhar99} Thus, 
\begin{equation}
\frac{{\partial}{\phi}}{{\partial}x}|_{x=0,L} = \frac{2{\pi}(2{\lambda_L}+d)[H_{dc}{\pm}H_{0}\sin({\omega}t)]}{\Phi_0},
\label{eq4}
\end{equation}
where $H_{dc}$ is the applied dc magnetic field, $H_0$ is the
amplitude of the microwave magnetic field at the edges of the junction, and
$\omega$ is the angular frequency of the microwave signal. The $\pm$ sign in
Eq.~(\ref{eq4}) indicates that the directions of the microwave
fields are opposite at the two edges of the junction. For this geometry, the
microwave electric field is in the y-direction which is defined to be
normal to the junction area, and is given by
\begin{equation}
E_{y}(x,t) = \frac{\Phi_0}{2{\pi}d}
\frac{{\partial}{\phi(x,t)}}{{\partial}t}.   
\end{equation}
The impedance and harmonic generation can then be
calculated from the Fourier transform of the time-dependent electric field
$E_y$,\cite{clem97}   
\begin{equation}
{R_n} = \frac{2}{H_0}{{\int}_{0}}^{T_{\rm rf}}dt{E_{y}(0,t)}{\sin(n{\omega}t)},
\label{eq6}
\end{equation}
\begin{equation}
{X_n} = \frac{2}{H_0}{{\int}_{0}}^{T_{\rm rf}}dt{E_{y}(0,t)}{\cos(n{\omega}t)},
\label{eq7}
\end{equation} 
where $T_{\rm rf}$ is the microwave period, $n$ is a positive integer, $R_1$ and
$X_1$ are proportional to the microwave resistance and
reactance at the fundamental resonator frequency, and $R_n$ and $X_n$
for $n>1$ correspond to the $n$th harmonic generated in the
junction. In Eqs. (6) and (7) we use only $E_y$ at the edge of the 
junction, since the microwave loss is dominated by the behavior at the
edges as found by Lehner $et$ $al$.\cite{oates99} The calculated
resistance is compared with the measured results as a function of
$H_{dc}$ at $T=5$\,K in Fig.~\ref{fig3}. The parameters used in the
calculation are ${J_c}=4.078\times{10^2}$ A/cm$^2$,
$\rho_n=6.75\times{10^{-8}}\,{\Omega}$\,cm$^2$, where $\rho_n$ is obtained from dc I-V
measurements, and $J_c$ is taken to fit the experimentally observed
$H_{cJ}$ (Eq. (2)). The calculation shows peaks in the microwave loss at
the same dc magnetic fields as the experimental results. 

We interpret the peaks in $1/{Q_0}$ as the result of Josephson vortices entering
the junction. After zero-field cooling, the applied dc magnetic field
is gradually increased and as long as $H_{dc} <H_{cJ}$, the external
magnetic field is screened by the self-current in the junction
(Meissner state). A further increase of the applied dc magnetic field
so that $H_{dc}{\approx}H_{cJ}$ causes a Josephson vortex to almost
enter the junction. Since the junction has two edges, this event is
actually a two-vortex event, one from each edge. Because of the
presence of the small rf magnetic field at the junction edges, two
Josephson vortices are created and annihilated during each rf
cycle. This state of the junction manifests itself in our experiment
with sharply decreased $Q_0$ because the power dissipation is the
highest when a Josephson vortex is created or annihilated at the
edges.\cite{oates99} With a slightly higher applied dc field, the
Josephson vortex is driven completely into the junction and there is
very little extra microwave loss once the vortex is in the
junction.\cite{oates99} As the applied field is increased further, the process 
is repeated each time another vortex enters the junction, until the
junction is packed with Josephson vortices, and the collective
properties of a large number of strongly interacting vortices have to be
considered. Our numerical results are also qualitatively consistent
with a previous linearized analytical treatment of the microwave
absorption by a long junction in a magnetic field.\cite{koshelev89}

The temperature dependence of $H_{cJ}$ comes from that
of $J_c$ and $\lambda_L$ in accordance with Eq.~(\ref{eq1}). The
experimental temperature dependence of the onset of the first peak in
the microwave resistance agrees roughly with the theoretically
predicted $H_{cJ}(T)$.\cite{barone} This
agreement is consistent with the hypothesis that the peaks observed in
the microwave resistance of the grain-boundary junction are caused by
single Josephson vortices entering the junction.

Since the calculation involving the rf magnetic field is computer-time
consuming, we have also solved the static sine-Gordon equation without
the perturbation from the rf signal. In order to get a physically
meaningful solution, we keep the damping term  $d{\phi}/dt$ in
Eq~(\ref{eq2}). After turning on the dc field at $t=0$, $\phi(x,t)$ is
integrated in time until a static solution is reached. Magnetic field
and current distributions in the junction are then obtained.  
We have confirmed that
the dc fields at which the microwave loss peaks appear 
correspond to the calculated dc fields
at which individual Josephson vortices enter the
junction. Therefore, results from the static dc calculation should
provide sufficient information about the microwave losses. 

We have 
calculated the hysteresis effects of the dc magnetic field  using
the static sine-Gordon solutions. To simulate our
experiment, we have carried out the calculation with a dc field increasing
from 0 to 3\,Oe, then decreasing from 3 back to 0\,Oe, with
each step using $phi(x)$ obtained from the previous step as the
initial condition. The results show that as the
dc field decreases, 
the Josephson vortices leave the junction at different field values than
those at which vortices enter while the dc field is
increasing. In addition, the number of expected microwave loss peaks (six)
is more than in the case of the increasing field (four) because at
some fields, one-half of a flux quantum leaves the junction, while all
fluxon entry events cause a change of one flux quantum as the field
increases.  The measured $1/{Q_0}$ versus $H_{dc}$ for decreasing $H_{dc}$ is
plotted in Fig.~\ref{fig4}. The
calculation-predicted dc fields at which vortices leave the junction
are indicated by the arrows in Fig.~\ref{fig4}. 
Peaks of the microwave loss are observed
but the agreement between the measured and the predicted results is
not as good as the case of the increasing field. The calculations have
considered only the case of a uniform junction. The good agreement
with  
the experimental data for an increasing field indicates that the
junction is sufficiently homogeneous for the assumption of uniform
$J_c$ to be a good approximation. The double-peak feature apparent in the measured
data in Fig.~\ref{fig3} might be due to a slightly asymmetric $J_c$
near the two edges of the junction so that there is one vortex
penetrating into the junction from each edge at slightly different
fields.  

We believe that nonuniform $J_c$ and the pinning resulting from it
might explain the differences seen in Fig.~\ref{fig4} between the
calculated and measured behavior in decreasing field. In the case of
increasing field, pinning is expected to have little effect on the
process of vortex entry, and once in the junction, the losses are low,
so the effects of pinning are not observed. On the other hand, pinning
would affect the fields at which the vortices leave the junction by
effectively adding to the potential barrier that a vortex must
overcome to exit the junction.
This is probably why the data for
increasing field agrees better with the calculation. It will be
interesting to consider defects in the junction in the calculation, so
that the pinning of the Josephson vortices can be included.  

The results of the measurements and calculations presented above show
that we are able to probe individual Josephson vortices 
entering and exiting the bicrystal grain-boundary Josephson junctions under
study. The microwave loss is not a monotonic function of external dc
magnetic field and increases dramatically when a single vortex
penetrates into or exits a long junction. Hysteretic behavior has been
observed both in the experiments and the calculation. Since it is believed that HTS films
contain weak links and the threshold field for vortex penetration can
be very small, even the earth's ambient field or the field from trapped
flux might have a significant impact on the microwave impedance of HTS
films. In addition, the collective effects from many Josephson
junction-like weak links might also explain the anomalous dc response
observed\cite{sridhar97} in which $R_S$ decreases with the application of a small magnetic field.


We gratefully acknowledge support for this work by the Air Force
Office of Scientific Research. The authors wish to thank L.~R. Vale
and R.~H. Ono at NIST Boulder, CO for providing the samples used in
this study, and J.~Derov, G.~Roberts, R.~Webster and J.~Moulton at
AFRL for their hospitality. We also wish to thank Dr B. Willemsen for
providing a rf probe used in this project.


\begin{figure}[tb]
\caption{The upper figure shows an engineered grain-boundary junction and
the patterned YBCO microstrip resonator used in this study. The first
($n$=1, $f_1$=4.4\,GHz) mode has a current peak at the junction, while
the second mode ($n$=2, $f_2$=8.7\,GHz) has a current node at the junction.}
\label{fig1}
\end{figure}

\begin{figure}
\caption{Measured $1/{Q_0}$ of the microstrip resonator as a function of
dc magnetic field at different temperatures. The threshold field at
which the first peak in the microwave loss appears can be identified
as the critical Josephson field $H_{cJ}$.}
\label{fig2}
\end{figure}

\begin{figure}[tb]
\caption{(a) Measured $1/{Q_0}$ at 5\,K vs increasing dc magnetic field
compared with (b) the calculated effective Josephson junction resistance
vs dc magnetic field. The series of peaks observed are identified
with individual Josephson vortices entering the junction.}
\label{fig3}
\end{figure}


\begin{figure}
\caption{Measured $1/{Q_0}$ at 5\,K vs dc magnetic field (decreasing). The
arrowes in the figure indicate the calculated fields at which a
quantized amount of flux leaves the junction.}
\label{fig4}
\end{figure}

\narrowtext

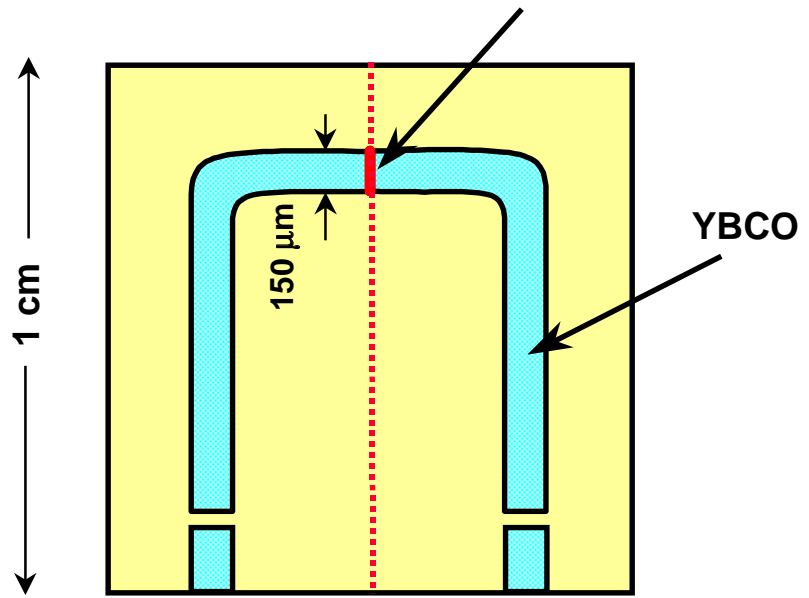
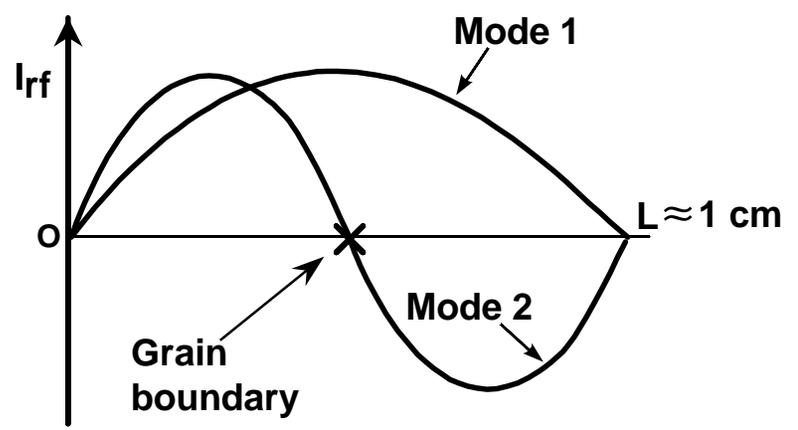
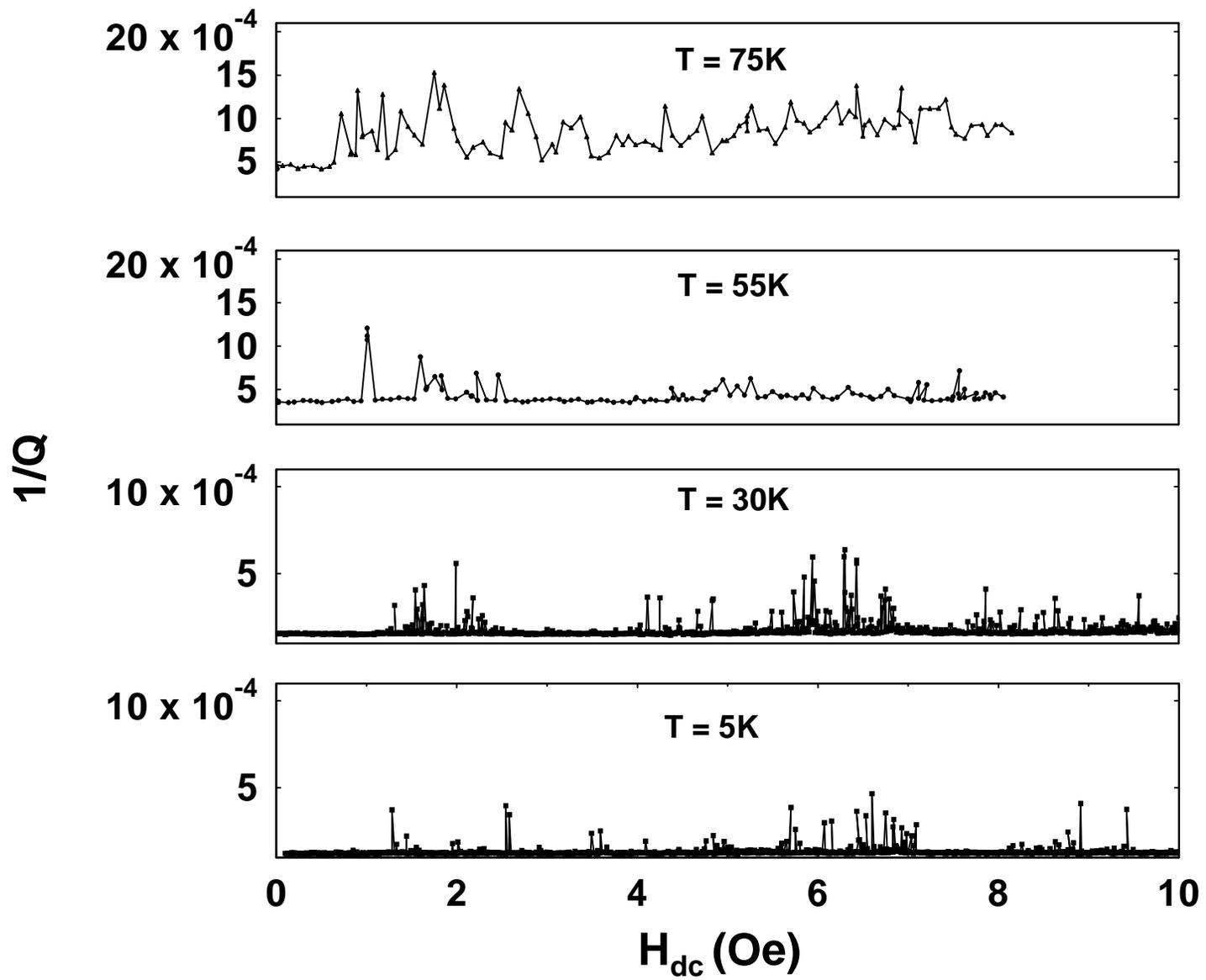
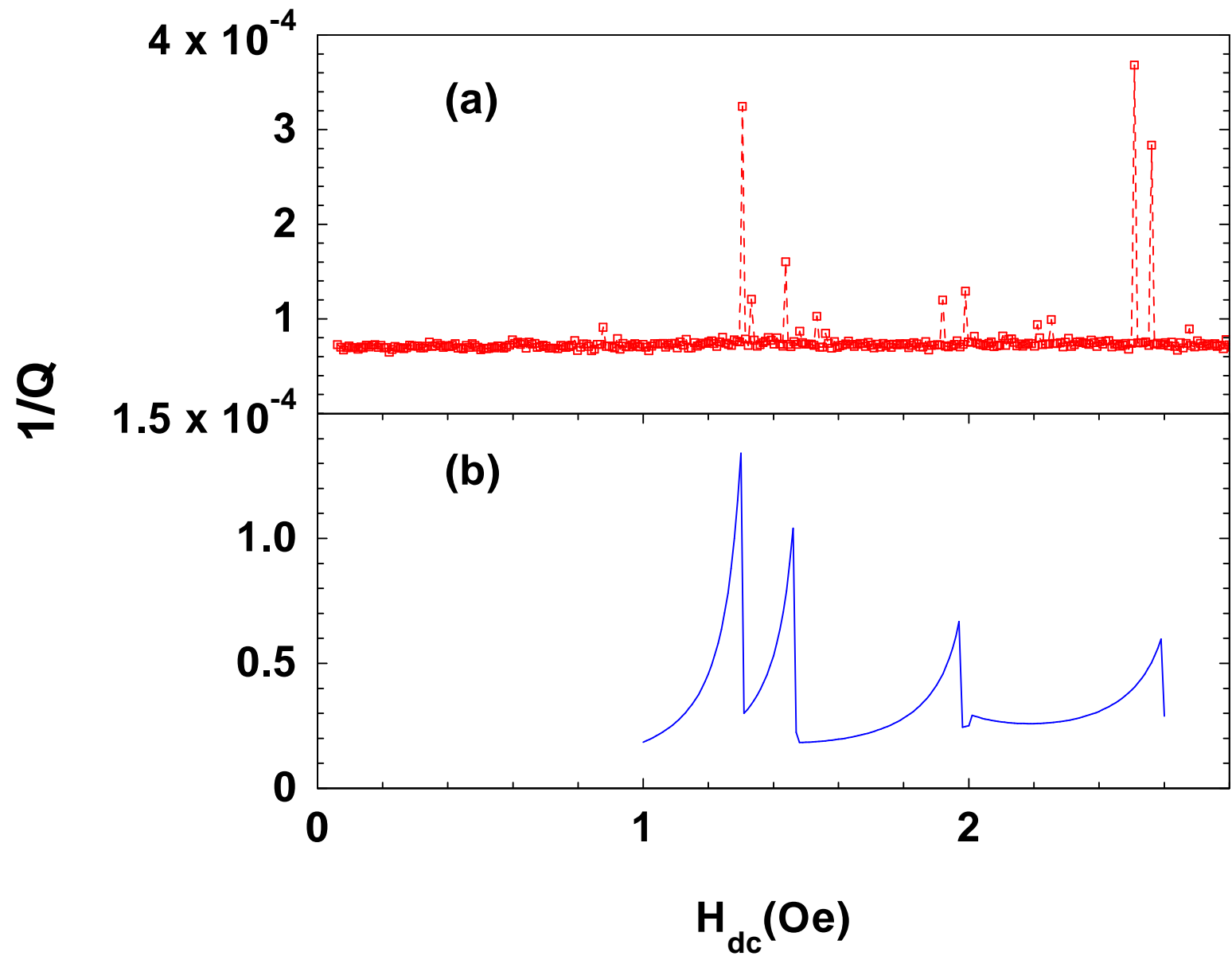
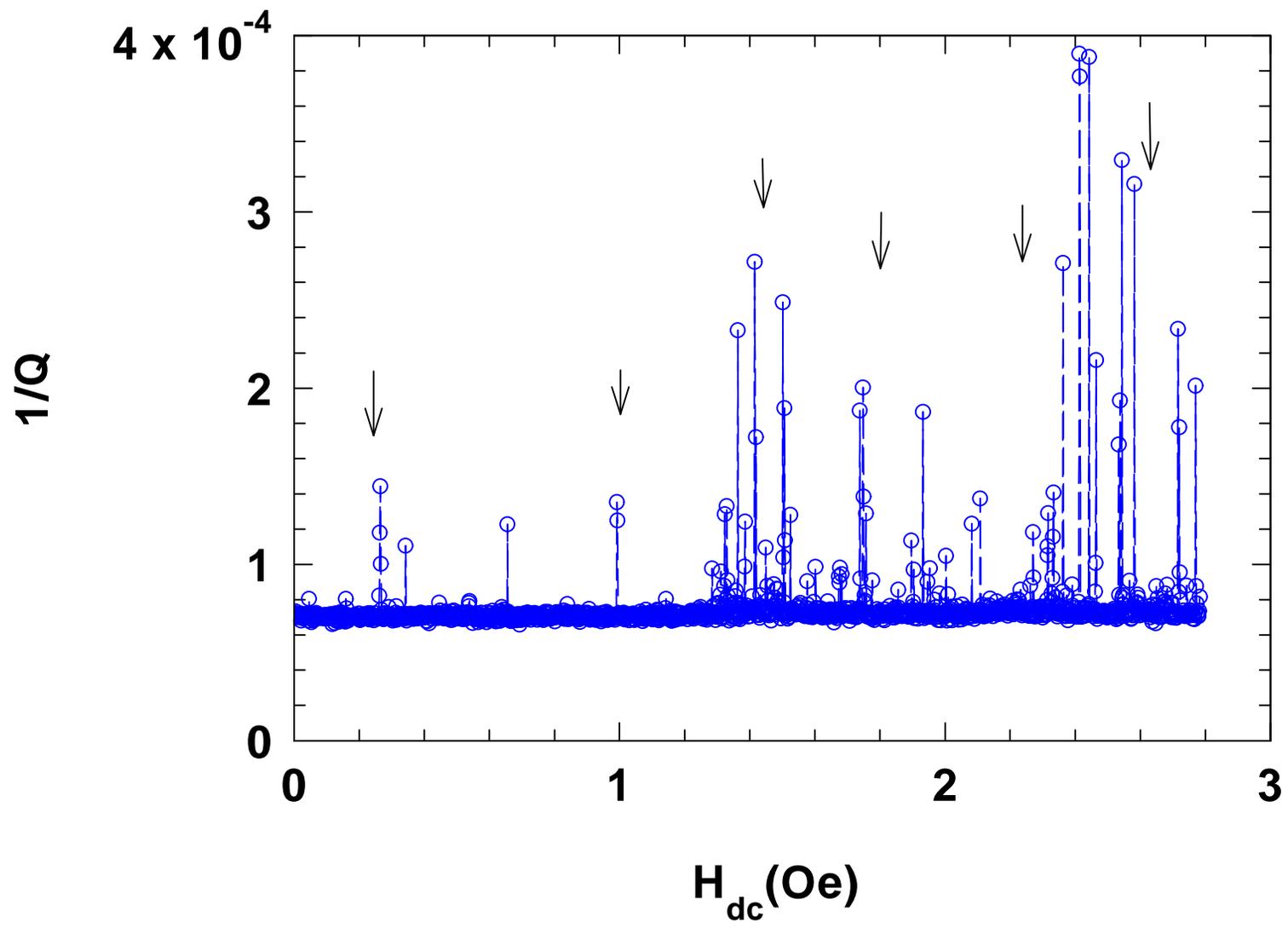
\end{document}